\documentclass[prl,showpacs,twocolumn,preprintnumbers,amssymb,amsmath]{revtex4}
\usepackage{graphicx}

\begin{document}

\bibliographystyle{apsrev}

\title{Non-extensive thermodynamics of 1D systems with long-range interaction}

\author{S. S. Apostolov$^{1,2}$, Z. A.
Mayzelis$^{1,2}$, O. V. Usatenko$^1$
\footnote[1]{usatenko@ire.kharkov.ua}, V. A. Yampol'skii$^1$}
\affiliation{$^1$ A. Ya. Usikov Institute for Radiophysics and
Electronics \\
Ukrainian Academy of Science, 12 Proskura Street, 61085 Kharkov,
Ukraine \\ $^2$ V. N. Karazin Kharkov National University, 4
Svoboda Sq., 61077 Kharkov, Ukraine}

\begin{abstract}
A new approach to non-extensive thermodynamical systems with
non-additive energy and entropy is proposed. The main idea of the
paper is based on the statistical matching of the thermodynamical
systems with the additive multi-step Markov chains. This general
approach is applied to the Ising spin chain with long-range
interaction between its elements. The asymptotical expressions for
the energy and entropy of the system are derived for the limiting
case of weak interaction. These thermodynamical quantities are
found to be non-proportional to the length of the system (number
of its particle).
\end{abstract}
\pacs{05.40.-a, 02.50.Ga, 05.50.+q}

\maketitle

One of the basic postulates of the classical statistical physics is
an assumption about the particle's interaction range considered to
be small as compared with the system size. If this condition does
not hold the internal and free energies, entropy, etc. are no more
additive physical quantities. Due to this fact the definitions of
the temperature, entropy, etc. are not evident. The distribution
function of the non-extensive system is not the Gibbs function, the
Boltzman relationship between the entropy and the statistical weight
is not any longer valid.

The non-extensive systems are common in physics \cite{daux}. They
are gravitational forces \cite{padma}, Coulomb forces in globally
charged systems \cite{nichol}, wave-particle interactions, magnets
with dipolar interactions \cite{barre}, etc. Long-range correlated
systems are intensively studied not only in physics, but also in the
theory of dynamical systems and the theory of probability. The
numerous non-Gibbs distributions are found in various sciences,
e.g., Zipf distribution in linguistic~\cite{zipf,kant}, the
distribution of nucleotides in DNA sequences and computer
codes~\cite{DNA,mbgh,vossDNA}, the distributions in financial
markets~\cite{par,stan,mant}, in sociology, physiology, seismology,
and many other sciences~\cite{sorn,http}.

The important model of the non-extensive systems in physics is the
\emph{Ising spin chain} with elements interacting at great
distances \cite{isi,cas}. Unfortunately, the generalization of the
standard thermodynamic methods for the case of arbitrary number of
interacting particles is impossible. There exist solutions for
some particular cases of particles interaction
only~\cite{theodor,isi}. The other results obtained for the chains
with long-range interaction are either the general theorems about
the existence of the phase transitions in the system or the
calculations of the critical indexes~\cite{kad,halp}.

Several algorithms for calculating the thermodynamic quantities of
the long-range correlated systems have been proposed.
Unfortunately, they are not well grounded and need additional
justifications. One of them is the \emph{Tsalis thermodynamics}
(see \cite{tsal,abe}) based on axiomatically introduced entropy: $
S^{(q)}(W)=(W^{1-q}-1)/(1-q)$. Here $W$ is the statistical weight
and $q$ is the parameter reflecting the non-extensiveness of the
system. However, this expression does not correspond to the
entropy of Ising chain with a long but finite range of
interaction. Indeed, it does not contain the size of the system
and remains non-additive when the length of the system tends to
infinity. Meanwhile, the entropy has to be additive if the system
is much greater than the characteristic range of interaction.
Thus, the problem of finding the thermodynamic quantities for the
non-extensive systems is still of great importance.

Recently a new vision on the long-range correlated systems, based on
the association of the latter with the \emph{Markov chains}, has
been proposed~\cite{uya}. The binary Markov chains are the sequences
of symbols taking on two values, for example, $\pm 1$. These
sequences are determined by the conditional probability of some
symbol to occur after the definite previous $N$ symbols and this
conditional probability does not depend on the values of symbols at
a distance, greater than $N$, $P(s_i=s|T^-_{i,\infty}) =
P(s_i=s|T^-_{i,N})$. Here $T^\pm_{i,L}$ are the sets of $L$
sequential symbols $(s_{i\pm 1},s_{i\pm 2},\dots,s_{i\pm L})$. The
unbiased \emph{additive} Markov chain is defined by the additive
\emph{conditional probability function},
\begin{equation}\label{adm}
P(s_i=1 \mid T^-_{i,\infty}) =\dfrac{1}{2}+\sum_{r=1}^{N}
F(r)s_{i-r}.
\end{equation}
The function $F(r)$ is referred to as the memory
function~\cite{muyag}. It describes the direct interaction between
the elements of the chain contrary to the binary correlation
function $K(r)=\overline{s_i s_{i+r}}$ that also takes into
account the indirect interactions. Here
$\overline{\phantom{Z}\!\!\!\dots}$ means statistical averaging
over the chain. As was shown in~\cite{MUYa05}, the correlation
function can be found from the recurrence relation
$K(r)=\sum_{r'=1}^{N} 2F(r')K(r-r')$, that establishes the
one-to-one correspondence between these two functions. In the
limiting case of small $F(r)$, this equation yields $K(r)\approx
\delta_{r,0}+ 2F(r)$.

We consider the \emph{chain of classical spins} with hamiltonian
\begin{equation}\label{hamilt}
\mathrm{H}=-\sum\limits_{j-i<N\atop i<j}\varepsilon(j-i)s_i s_j,
\end{equation}
where $s_i$ is the spin variable taking on two values, $-1$ and $1$,
and $\varepsilon(r)>0$ is the exchange integral of the ferromagnetic
coupling. The range $N$ of spin interaction may be arbitrary but
finite.

We find thermodynamical quantities, specifically, energy and
entropy, of such a chain being in thermodynamical equilibrium with a
Gibbs thermostat of temperature $T$. The hamiltonian~\eqref{hamilt}
does not include the terms, corresponding to the interaction of the
system with the thermostat. Nevertheless, this interaction leads to
the relaxation of the temperature in different parts of the chain,
and the thermostat fixes its temperature of the whole spin chain. A
great many numerical procedures, for example, the Metropolis
algorithm, were elaborated for achieving the equilibrium state. The
algorithm consists in the sequential trials to change the value of a
randomly chosen spin. The probability of spin-flip is determined by
the values of spins on \emph{both sides} of the chain within the
interaction range. Thus, the Ising spin chain in the equilibrium
state can be characterized by the conditional probability, $P(s_i=s
\mid T^-_{i,\infty},T^+_{i,\infty})$, of some spin $s_i$ to be equal
to $s$ under the condition  of definite values of all other spins on
\emph{both sides} from $s_i$. In Ref.~\cite{equiv}, a chain with the
two-sided conditional probability function, which is independent of
symbols at a distance of more than $N$, was considered. These chains
were shown to be equivalent to the $N$-step Markov chain. So, to
calculate the statistical properties of the physical object with
long-range interaction, it is sufficient to find the corresponding
Markov chain.

In this work, we demonstrate that the Ising spin chains with
long-range interaction being in the equilibrium state are
statistically equivalent to the Markov chains with some
conditional probability function. Then, using the known
statistical properties of the Markov chains, we calculate the
thermodynamical quantities of the corresponding spin chains.

First, we present a method of defining a thermodynamical quantity
$Q$ (e.g., energy, entropy) for a subsystem of arbitrary length
$L$ (a set of $L$ sequential particles) of any non-extensive
system. This subsystem, denoted by $S$, interacts with the
thermostat and with the rest of the system. The interaction with
the latter makes it impossible to use the standard methods of
statistical physics. The internal energy of the entire system is
not equal to the sum of the internal energies of $S$ and the rest
of the system.

In order to find the condition of the system equilibrium we divide
the ensemble of the system into the \emph{subensembles} with fixed
$2N$ closest particles from both sides of $S$. We denote these two
''border'' subsystems of length $N$ by letter $B$. All the other
particles, except for $S$ and $B$, are denoted by $R$:
\begin{widetext}
\[
\underbrace{\dots s_{i-N}}\limits_{R}\underbrace{s_{i-N+1} \dots
s_{i}}\limits_{B} \underbrace{s_{i+1} \dots
s_{i+L}}\limits_{S}\underbrace{s_{i+L+1}\dots
s_{i+L+N}}\limits_{B}\underbrace{s_{i+L+N+1}\dots}\limits_{R} \]
\end{widetext}
Within every subensemble, the subsystem $B$ is fixed and plays the
role of a partition wall conducting the energy and keeps its
internal energy unchanged. The energy of system $S+R$ equals to the
sum of energies of $S$ and $R$ and their energy of interaction with
$B$. Thus, within every subensemble we can use the equilibrium
condition between $S$ and $R$, analogous to that for extensive
thermodynamics:
\begin{equation}\label{equil_nonext}
\dfrac{\partial \ln W_S(E_S|B)}{\partial E_S}=\dfrac{\partial \ln
W_R(E_R|B)}{\partial E_R},
\end{equation}
where $W_S(E_S|B)$ and $W_R(E_R|B)$ are the statistical weights of
systems, in which the energy of $S$ is $E_S$, those of $R$ is
$E_R$, and $B$ is fixed. We refer to these statistical weights as
the \emph{conditional statistical weights}.

In a similar way, we introduce any conditional thermodynamical
quantity $Q(\cdot|B)$.
The real quantity $Q(\cdot)$ is the conditional one, averaged over
the subensembles with different environments $B$:
\begin{equation}
Q(\cdot)=\big\langle Q(\cdot|B)\big\rangle_{B}
\end{equation}
Using this way of calculating the thermodynamical quantities, one
does not need to find the distribution function of the system $S$.
Nevertheless, it is quite appropriate for the non-extensive
systems and yields the thermodynamical values that could be
measured experimentally. If the system is in the thermal contact
with the Gibbs thermostat, within every subensemble with a fixed
$B$, the equilibrium condition between thermostat and subsystem
$S$ yields the temperature of $S$ equal to $T$. Thus, the averaged
temperature of the subsystem $S$ is also $T$.

The conditional entropy can be introduced as the logarithm of the
conditional statistical weight: $S(E_S|B)=\ln W(E_S,S|B)$. Equality
$dS(E_S|B)=dE_S/T$ is fulfilled for the conditional quantities $S$
and $E$. Meanwhile, such a relation is not valid for the averaged
entropy and energy.

Note that the presented algorithm for calculating the
thermodynamical quantities is rather general and can be applied to
other quantities, e.g., to the probability for some spin to take on
the definite value under the condition of the particular
environment.

Now we proceed to the application of this algorithm for the
analysis of \emph{Ising spin chain} with long-range interaction.
The theory of Markov chains is built around the expression for the
\emph{conditional probability function}. So, in an effort to find
a Markov chain that corresponds to a given Ising system, we have
to find its conditional probability function. To this end, we
consider one spin $s_i$ as a system $S$ and the conditional
probability $P$ as a quantity $Q$ in the above-mentioned
algorithm. In the extensive thermodynamics, this probability is
determined by the Gibbs distribution function. According to our
algorithm, one can find that the probability of event $s_i=1$
under the condition of the fixed spins from $B$ is given as
\begin{equation}\label{uslver}
p(s_i=1|B)=\Big(1+\mathrm{exp}\big(-\sum \limits
_{r=1}^{N}\dfrac{2\varepsilon
(r)}{T}(s_{i-r}+s_{i+r})\big)\Big)^{-1}.
\end{equation}

It should be pointed out that this expression is similar to the
Glauber formula~\cite{glaub} written in some other terms. Our
result has been that the conditional probability of the $s_i$
occurring is determined by two subsystems of length $N$ on both
sides of $s_i$ only and does not depend on the remoter spins. As
mentioned above, this is a property of the $N$-step Markov
sequences.

To arrive at an explicit expressions for the \emph{energy and
entropy} of the subsystem $S$ we consider the case of \emph{weak
interaction}, $\sum \limits _{r=1}^{N}\varepsilon (r)\ll T$. In
this case, Eq.~\eqref{uslver} corresponds approximately to the
additive Markov chain with the memory function $F(r)\approx
\varepsilon (r)/2T$ and, thus, its binary correlation function is
\begin{equation}\label{coris}
K(r)\approx \delta_{r,0}+\varepsilon (r)/T.
\end{equation}

\begin{figure}[h!]
\begin{centering}
\scalebox{0.8}[0.8]{\includegraphics{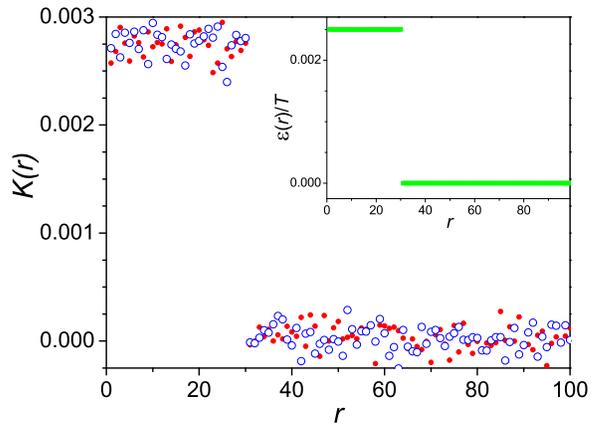}} \caption{(Color
online) The correlation functions of the additive Markov chain
determined by memory function $F(r)=\varepsilon (r)/2T$ (solid
circles) and the spin chain being equilibrated by the Metropolis
scheme (open circles). The inset represents the function
$\varepsilon (r)/T$.} \label{corr}
\end{centering}
\end{figure}

Figure \ref{corr} shows that the Ising spin chain being in
equilibrium state is statistically equivalent to the Markov chain.
The solid circles describe the binary correlation function of the
additive Markov chain with the memory function $F(r)=\varepsilon
(r)/2T$. The open circles correspond to the correlation function
of the spin chain being equilibrated by the Metropolis scheme.

It can be shown that the Metropolis scheme also yields the same
expression~\eqref{uslver} for the conditional probability. Due to
this fact one does not need to use the Metropolis algorithm to find
an equilibrium state of the spin chain but can generate a Markov
chain with a corresponding conditional probability function.

Now we examine \emph{binary spin chain} determined by the
Hamiltonian~\eqref{hamilt} with $s_{i}=\pm 1$ and find the
non-extensive energy and entropy of subsystem $S$ containing $L$
particles. Since the explicit expressions for the correlation
function were derived for the additive Markov chains with the
small memory function only, we suppose the energy of interaction
to be small as compared to the temperature,
$\sum\limits_{r=1}^N\varepsilon(r) \ll T$.

The proposed algorithm of finding thermodynamic quantities is
considerably simpler while finding the \emph{energy} of the system
$S$ of length $L$. The energy is the averaged sum of products of
pairs of spin values,
\begin{equation}\label{eee}
E(T)=\overline{\sum_{j=i}^{i+N}\sum_{k} \varepsilon(|k-j|)s_k s_j
-\sum_{j,k=i}^{i+N} \varepsilon(|k-j|)s_k s_j}.
\end{equation}
Formally, this expression depends on the index $i$ of the first
spin in the system $S$. However, we study the homogeneous systems,
so function $E(T)$ in Eq.~\eqref{eee} does not contain $i$ as an
argument. Energy in Eq.~\eqref{eee} can be calculated via the
binary correlation function only with the arguments, less than
$N$, without finding conditional energies. Using
Eq.~\eqref{coris}, we arrive at
\begin{equation}
E(T)=-\Big(\epsilon^2_{N} L+\sum\limits_{i=1}^N \varepsilon^2(i)
\min \{i,L \}\Big)/T, \label{energy}
\end{equation}
with $\epsilon^2_{N}=\sum\limits_{i=1}^N \varepsilon^2(i)$. If the
system length is much greater than memory length $N$ this
expression yields extensive energy. Indeed, in this case
subsystems $S$ and $R$ interact nearly extensively. In the
opposite limiting case we get:
\begin{equation}
E(T) \approx -(4\epsilon^2_{N} L-\varepsilon^2(1)L^2)/2T, \quad L\ll
N.
\end{equation}
If one regards the \emph{whole} chain of the length $M$, forming the
circle, as the system $S$ in the above-mentioned sense, the additive
energy is
\begin{equation}\label{enlast}
E(T)=-\epsilon^2_{N} M/T.
\end{equation}
This result is very natural because the extensive thermodynamics is
valid.

\begin{figure}[t]
\begin{centering}
\scalebox{0.8}[0.8]{\includegraphics{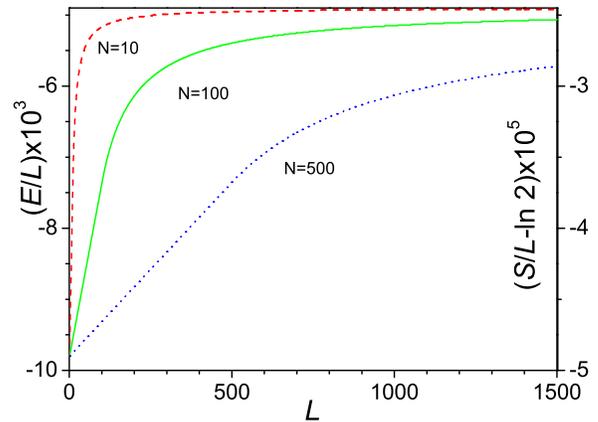}} \caption{(Color
online) The specific non-extensive energy $E\times10^{3}/L$ and
entropy $(S/L-\ln2)\times10^5$ vs. the size $L$ of system  for
step-wise interaction $\varepsilon(r)$. The memory length $N$ is
indicated near the curves. The constant limiting value of energy is
$-4\epsilon^2_{N}/2T\approx -5\times10^{-3}$.} \label{esht}
\end{centering}
\end{figure}

It is seen, that the non-extensive energy is expressed in terms of
binary correlation functions only. This is not correct for other
thermodynamical quantities, e.g. the \emph{entropy} of the system
$S$. Formally, to find the entropy, we have to calculate all
conditional entropies by \emph{integration} of
$dS(E|B)=dE(S|B)/T$, and then \emph{to average} the result over
all realizations of the borders. However, at high temperatures, to
a first approximation in the small parameter
$\sum\limits_{r=1}^N\varepsilon(r)/T$ we can change the order of
these operations and calculate the averaged entropy by integrating
this formula taken with the averaged energy. A constant of
integration is such that the chain is completely randomized at $T
\rightarrow \infty$, and its entropy is equal to $\ln 2^L$. Thus,
we obtain
\begin{equation}\label{entropy}
S(T)=L \ln 2+E(T)/2T.
\end{equation}
Here $E(T)$ is determined by one of
Eqs.~(\ref{energy})-(\ref{enlast}). This expression describes the
non-extensive entropy of the system $S$. However, while the energy
is non-extensive in the main approximation, the entropy is
non-extensive as a first approximation in the parameter
$\sum\limits_{r=1}^N\varepsilon(r)/T$. The dependences of the
non-extensive energy and entropy on size $L$ of the system $S$ are
given in Fig.~\ref{esht} for step-wise interaction $\varepsilon(r)$.
The solid line corresponds to additive quantities, it is the
asymptotic of these quantities for the large system length.

It should be emphasized, that knowing the energy and entropy we can
find some other thermodynamic quantities. For example, at high
temperatures the heat capacity can be determined in a similar way as
for the entropy calculating. One can use the classical formula
$C_V(T)=TdS(T)/dT$ with averaged entropy from Eq.~\eqref{entropy}
and obtain the heat capacity value as $C_V=-E(T)/T$. This simple
procedure of its finding is valid for the case of high temperatures
as the main approximation only. In the general case, $C_V(T)$ is not
determined by $C_V(T)=TdS(T)/dT$. This relation holds for the
conditional quantities only.

At high temperatures, we calculated the averaged non-additive
energy and entropy without using the conditional ones. In the
opposite limiting case of low temperatures, the calculation of
conditional quantities proves to be necessary.

Thus, we have suggested the algorithm of evaluating thermodynamic
characteristics of non-extensive systems. The value of certain
thermodynamical quantity can be obtained by averaging the
corresponding conditional quantity. This method is applied to the
Ising spin chain. The explicit expressions for the non-additive
energy and entropy are deduced in the limiting case of high
temperatures as compared to the energy of spins interaction. At high
temperatures, the equilibrium Ising chain of spin turns out to be
equivalent to the additive multi-step Markov chain.

\end{document}